\documentclass[11pt]{revtex4}
\usepackage{epsfig}
\usepackage{graphics}
\usepackage{amsmath}

\begin{document}
\title{Spontaneous polarisation of the neutral interface
for valence asymmetric coulombic systems.}
\author{D. di Caprio \footnote{E-mail: dung.dicaprio@gmail.com}}
\affiliation{Laboratory of Electrochemistry and Analytical Chemistry,\\
University Paris 6, CNRS, ENSCP,\\
Universit\'e Paris 6, B.P. 39\\
4, Place Jussieu, \ 75252 Paris Cedex 05, France}
\author{M. Holovko}
\affiliation{Institute for Condensed Matter Physics,
National Academy of Sciences\\1 Svientsitskii Str., 79011 Lviv, Ukraine\\}

\begin{abstract}
In this paper, we discuss the phenomenon of a spontaneous polarisation of a
neutral hard planar interface for valence asymmetric coulombic systems.
Within a field theoretical description, we account for the existence
of non trivial charge density and electric potential profiles.
The analysis of the phenomenon shows that the effect is related to combinatorics
in relation with the existence of the two independent species cations
and anions. This simple and basic feature is related to the quantum mechanical
properties of the system.
The theoretical results are compared with numerical simulations data and are
shown to be in very good agreement, which a fortiori justifies our physical
interpretation.
\end{abstract}

\maketitle

\setlength{\fboxrule}{0.8pt}
\section{Introduction}

There has been recently a renewed interest in the study of the structure
of the double layer.
This stems both from the existence of new important domains of application as
for instance those related with the nanotechnologies,
microfluidics, microbatteries, and electrochemical sensors, biology or
also the use of new ionic liquid solvents \cite{ajdari,kornyshev}.
These new domains represent a theoretical challenge as the behaviours of the
charged systems does not in general follow the popular predictions of the
Poisson-Boltzmann approach.
There can be many reasons for parting from the standard behaviour. For instance,
it can be related to excluded volume like packing effects at high
local concentrations at the interface \cite{ajdari,kornyshev} but also, in
the regime of low reduced temperatures, to the existence of strong Coulomb
interactions which require to account for correlations beyond mean field
theories like the popular Gouy-Chapman (GC) approach.
In the latter case, there has been a series of papers
\cite{BodaChan,BodaChan2,MierYTeran,BodaHolovko,Pizio,BodaSokolowski,
Outhwaite1,Fawcettconj,anomalouscapa,bodaasym,Outhwaite2} devoted to the challenging
problem of the anomalous behaviour of the electric capacitance.

In this paper, we investigate another type of systems which also depart from the
standard GC approach. It is the case of the asymmetric in valence electrolytes.
In an early paper, Torrie et al. \cite{Torrie} have shown the existence of a
polarization of the interface even at the point of zero charge (PZC) for an
extended restricted primitive model of ions with different valences.
This phenomenon is far from obvious and standard intuitive approximations like
the GC theory are unable to describe it.
In contrast to the case of asymmetric in size ions, where the smallest ion by
coming closer to the interface induces a polarization of the interface, the
behaviour of valence asymmetric ions is rather non intuitive.
Torrie et al. have used a modified Poisson-Boltzmann approach to describe this
effect which is in our opinion rather costly in its mathematical application.
More recently Henderson et al. \cite{HendersonAsym} have addressed the problem
using standard approximations in the liquid state theory searching for
simple analytic expressions. They were able to reproduce the polarization
effect for the size asymmetry but not for the valence asymmetry.
From a practical point of view, the interest of studying the behaviour of the PZC
is that this quantity is often considered as an indication of chemical
interaction of the ions with the electrode so called "specific adsorption".
Here, by considering a hard wall, we discard the role of "specific interactions"
and intend to propose a simple physical interpretation.

The article is organized as follows.
In the first Section, we present the field theory formalism
for valence asymmetric electrolytes and introduce the meaningful
physical parameters for the system.
In the following two Sections \ref{sec:g_q} and \ref{sec:vpzc}, we derive
respectively the expressions of the charge profile and of the electric potential
accross this interface at a neutral hard wall. Then in Section
\ref{sec:chargecontact}, we compare the results to the exact relation given by
the charge contact theorem \cite{chargecontact} and verify the consistency
of our expressions.  Finally in Section
\ref{sec:discussion}, we discuss the different profiles with
respect to numerical simulations results \cite{HendersonAsym}.

\section{Field theory for point ions.}
\subsection{Formalism}
We consider a field theoretical description of a system of point ions situated
in a half space, in contact with a hard wall. The dielectric constant
$\varepsilon = \varepsilon_r \varepsilon_0$ is uniform throughout the space.
As in \cite{jphysa,anomalouscapa},
we introduce a field theoretical expression of the grand potential in terms of
the fields $\rho_+ (\mathbf{r})$ and $\rho_- (\mathbf{r})$ representing in space
the density fields for cation and anion distributions, respectively
\begin{eqnarray}\label{eq:PartFct0}
  \Theta[\rho_\pm ]= \int\mathcal{D}\rho_\pm (\mathbf{r})\exp\{-\beta H[\rho_\pm ] \}
\end{eqnarray}
where $\beta=1/(k_BT)$ is the inverse temperature.
The grand potential is then obtained as
$\beta (-p V + \gamma A) = -\ln \Theta[\rho_\pm]$, where $p$
is the pressure, $V$ is the volume, $\gamma$ is the surface tension,
and $A$ is the area of the electrodes.
For the study of the Coulomb interactions, it is convenient to use the charge
density field $q(\mathbf{r})=z_+ \rho_+(\mathbf{r}) - z_- \rho_-(\mathbf{r})$,
with $z_+$, $z_-$ the valences of cations and anions respectively,
and the total density field $s(\mathbf{r})=\rho_+(\mathbf{r}) + \rho_-(\mathbf{r})$.
The Hamiltonian as a functional of these fields is
\begin{equation}\label{eq:hamilt}
   \beta H  [q,s] =
    \beta H^{ent}[q,s]  +
    \beta H^{Coul}[q] - \int \beta\mu_s s(\mathbf{r}) d\mathbf{r}
    - \int \beta\mu_q q(\mathbf{r}) d\mathbf{r}
\end{equation}
where $\mu_s=\mu_++\mu_-$, $\mu_q=z_+\mu_+-z_-\mu_-$, and $\mu_\pm$ are the
chemical potentials of the ions.
The entropic part of the Hamiltonian which accounts for the quantum mechanical
degrees of freedom in the phase space \cite{jphysa} and the terms
related to the chemical potentials can be written as
\begin{eqnarray}
  \beta H^{ent}[q,s] - \int \beta\mu_s s(\mathbf{r}) d\mathbf{r}
    - \int \beta\mu_q q(\mathbf{r}) d\mathbf{r} &=&
    \int \frac{q({\bf r})+s({\bf r})}{2} \left[\ln
    \left(\frac{q({\bf r})+s({\bf r})}{2\bar{\rho}_+}\right) - 1\right] d{\bf r}\nonumber\\
    & &+ \int \frac{s({\bf r})-q({\bf r})}{2} \left[\ln
    \left(\frac{s({\bf r})-q({\bf r})}{2\bar{\rho}_-}\right) - 1\right] d{\bf r}
\end{eqnarray}
where $\bar{\rho}_\pm=\exp(-\beta \mu_\pm)/\Lambda^3$ with $\Lambda$ being the
de Broglie wavelength. The second term in the Hamiltonian is the coulombic
contribution
\begin{equation}
      \beta H^{Coul}[q] =
        \frac{\beta e^2}{8\pi\varepsilon} \int
        \frac{q({\bf r})q({\bf r'})}
        {|{\bf r}-{\bf r'}|}d{\bf r}d{\bf r'}
\end{equation}
where $e$ is the elementary electric charge.

In order to perform the calculations we expand the Hamiltonian of
equation~(\ref{eq:hamilt}) around the mean field profiles $\bar{q}=0$ and
$\bar{s}=\bar{\rho}$ which minimize the Hamiltonian in the case of the neutral
hard wall \cite{ddcjsjpbElectActa2003}, where $\bar{\rho}=\bar{\rho}_+ + \bar{\rho}_-$.
Beyond the mean field solution, introducing $\delta s=(s-\bar{\rho})/\bar{\rho}$
and $\delta q=q/\bar{\rho}$, the Hamiltonian is
\begin{eqnarray}
  \beta H &=& \bar{\rho}V
    + \frac{\bar{\rho}}{2}\int \left[\delta s^2(\mathbf{r}) +
    \frac{\delta q^2(\mathbf{r})}{z_{is}^2}\right] d\mathbf{r}
    +\beta \delta H + \beta H^{Coul} [q]
\end{eqnarray}
where $\beta \delta H$ contains terms of the expansion of orders higher than quadratic.
We introduce $z_{is}=\sqrt{z_+z_-}$ and $z_{as}=(z_+-z_-)/\sqrt{z_+z_-}$.
The first coefficient is related to the ionic strength as
  $z_+^2\bar{\rho}_++z_-^2\bar{\rho}_-=z_+z_-\bar{\rho}$.
However this parameter is not sufficient, as $z_{is}$ is the same for a 2:2 or a
4:1 electrolyte and it is the second coefficient $z_{as}$ which is really
characteristic of the asymmetry in valence between ions. With these notations we have
\begin{eqnarray}  \label{eq:deltaH}
  \beta \delta H &=& - \frac{\bar{\rho}}{2z_{is}^2}\int
    \left[ \delta s(\mathbf{r}) -\delta s^2(\mathbf{r})\right] \delta q^2(\mathbf{r}) d\mathbf{r}
    - \frac{z_{as}\bar{\rho}}{3! z_{is}^{3}}
      \int \left[ 1 - 2\delta s(\mathbf{r})+3\delta s^2(\mathbf{r}) \right]
      \delta q^3(\mathbf{r}) d\mathbf{r} \nonumber\\
    &&+ \frac{2\bar{\rho}}{4!z_{is}^4}\int (1+z_{as}^2)
   \left[ 1 - 3 \delta s(\mathbf{r})+6\delta s^2(\mathbf{r}) \right]
          \delta q^4(\mathbf{r}) d\mathbf{r} + ...
\end{eqnarray}
Note that the quadratic term in the Hamiltonian remains diagonal
for the $q$ and $s$ fields also for the valence asymmetric systems
and odd terms in $q$ appear only in $\beta \delta H$, the smallest odd
power in $q$ being three.

As in \cite{dungbodaasym}, the Hamiltonian can be further simplified by
scaling the charge density field $\delta q \rightarrow \delta Q=\delta
q/z_{is}$ and  by defining a new unit charge $\tilde{e}=z_{is}e$. In this case
for the even terms we recover the expansion of the symmetric 1:1 electrolyte and
the Hamiltonian is
\begin{eqnarray}
  \beta H &=& \bar{\rho}V
    + \frac{\bar{\rho}}{2}\int \left[\delta s^2(\mathbf{r}) +
    \delta Q^2(\mathbf{r})\right] d\mathbf{r}
    +\beta \delta H + \beta H^{Coul} [Q(\mathbf{r})]
\end{eqnarray}
with
\begin{eqnarray}\label{eq:deltaHren}
  \beta \delta H &=& - \frac{\bar{\rho}}{2}
   \int \delta s(\mathbf{r}) \delta Q^2(\mathbf{r}) d\mathbf{r}
   + \frac{\bar{\rho}}{12}\int \delta Q^4(\mathbf{r}) d\mathbf{r}
   - \frac{\bar{\rho}z_{as}}{6}\int \delta Q^3(\mathbf{r}) d\mathbf{r}
   + \frac{\bar{\rho} z_{as}^2}{12}\int \delta Q^4(\mathbf{r}) d\mathbf{r} + ...
\end{eqnarray}
where we have kept from equation~(\ref{eq:deltaH}), for clarity, only the terms
of interest. The Coulomb interaction term is also modified
\begin{equation}
      \beta H^{Coul}[q] =
        \frac{K_D^2}{8\pi\bar{\rho}} \int
        \frac{\delta Q({\bf r})\delta Q({\bf r'})}
        {|{\bf r}-{\bf r'}|}d{\bf r}d{\bf r'} ,
\end{equation}
where we have introduced the inverse Debye length,
$K_D =(\beta (z_+^2\bar{\rho}_++z_-^2\bar{\rho}_-) e^{2}/\varepsilon)^{1/2}
=(\beta \bar{\rho} \tilde{e}^{2}/\varepsilon)^{1/2}$,
which shows that $K_D$ has the same expression as for the 1:1 electrolyte except
that the elementary charge is now $\tilde{e}$.
Note that the scaling introduced in this section can be considered as a simple
renormalisation of the electrostatic quantities in relation with the ionic
strength.  In the following, we perform all calculations in terms of the charge
density field $\delta Q$ and transform the result in the physical field $\delta q$
at the end of the calculations.

The parameter $z_{is}$ has been completely absorbed in the new field
$\delta Q$ and the charge $\tilde{e}$ and like in \cite{dungbodaasym} we can use
the results for the 1:1 symmetric electrolyte and 
we mainly need to focus on the new terms in comparison to the symmetric
case which are associated with the asymmetry parameter $z_{as}$.
For the interfacial properties, at the one loop level of the calculations
we will assimilate the activity $\bar{\rho}$ and the average density $\rho$.

\section{Charge profile at the PZC}\label{sec:g_q}
\begin{figure}[h]
\begin{center} \epsfig{file=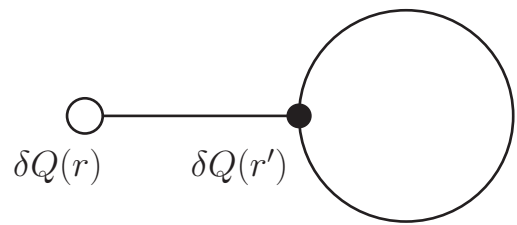,height=3.0cm,width=7.cm}
\end{center}
\caption{Diagram for the calculation of the charge density profile
as described in the text.
}\label{fig:diagram}
\end{figure}

The calculation of the charge profile illustrates the peculiarity of the
asymmetric systems.
For the symmetric $z:z$ electrolytes at the neutral interface, for reasons of
symmetry, average quantities which include an odd number of charge densities,
as for instance the charge profile, vanish.
Another way of stating this is by noting that it is impossible by applying the
Wick theorem to pair the charge fields and simultaneously calculate a quantity
with an odd number of charge fields, because this would imply the existence of
odd coupling constants.
Such odd terms exist for the asymmetric systems as a consequence of the
expansion of the entropic contribution to the Hamiltonian. The first
contribution is due to the three body coupling in equation (\ref{eq:deltaHren})
and corresponds to the diagram in figure \ref{fig:diagram} where we have the
standard $1/2$ symmetry coefficient associated and the analytic expression is
\begin{eqnarray}
  \langle \delta Q(\mathbf{r})\rangle =\frac{\rho z_{as}}{2}
   \int d\mathbf{r}' <\delta Q(\mathbf{r})\delta Q(\mathbf{r}')> <\delta Q(\mathbf{r}')\delta Q(\mathbf{r}')>
\end{eqnarray}
where the inhomogeneous correlation functions are those given in \cite{dungbodaasym},
where
\begin{eqnarray}\label{eq:qq_d}
  \left\langle \delta Q(\mathbf{r})\delta Q(\mathbf{r}')\right\rangle &=&
         \frac{1}{{\rho}}\left[\delta(\mathbf{r}-\mathbf{r}')
   - \frac{{K}_D^{2}}{4\pi}
        \frac{e^{-{K}_{D}|\mathbf{r}-\mathbf{r}'|}}{|\mathbf{r}-\mathbf{r}'|}
   +\int \frac{d\mathbf{K}}{(2\pi)^2}\;e^{-i\mathbf{K} (\mathbf{R}-\mathbf{R}') -K^{\prime}(x+x')}
   \frac{{K}_D^2\left(K-K^{\prime}\right)}
   {2K^{\prime}\left(K+K^{\prime}\right)}
   \right]\nonumber\\
\label{eq:qq_s}
\end{eqnarray}
where $x$, $x'$ are distances of the points from the wall and $\mathbf{R}$,
$\mathbf{R}'$ are respectively the projections of $\mathbf{r}$ and $\mathbf{r}'$
parallel to the wall and where $K^{\prime}=\sqrt{K^2+{K}_D^2}$.
In the previous correlation, when calculated at the same point, we renormalize
the dirac distribution as indicated in \cite{jphysa}. The bulk part which does not
depend on the distance to the wall, vanishes with the rest of the graph
as a consequence of the electroneutrality condition
$\int\left\langle\delta Q(\mathbf{r})\delta Q(\mathbf{r}')\right\rangle d\mathbf{r}'=0$.
And finally the excess contribution at the wall gives \cite{molphys1}
\begin{eqnarray}
  \left\langle \delta Q(\mathbf{r})\delta Q(\mathbf{r})\right\rangle&=&
\frac{K_D^3}{4\pi\bar{\rho}} I(\hat{x})
\end{eqnarray}
where we used the reduced distance to the wall $\hat{x}=x K_D$ and
 $I(\hat{x}) = \int_1^\infty { e^{-2\hat{x}t}}/{(t+\sqrt{t^2-1})^2} dt$.

We finally obtain our result, the profile for the charge
\begin{eqnarray}\label{eq:g_q}
  q(\hat{x}) = \frac{1}{2}z_{as}z_{is}\;\eta \rho\;F(\hat{x})
\end{eqnarray}
where the coefficient $\eta = K_D^3 / (8 \pi \rho)$ and the function
$F(\hat{x})=-2 I(\hat{x}) + f(\hat{x}) + c_0$
where
\begin{eqnarray}
  f(\hat{x}) = \int_1^\infty \left[\frac{e^{-\hat{x}}}{(2t-1)}
   -2 \frac{ e^{-2\hat{x}t}}{(2t-1)(2t+1)} \right]
   \frac{dt}{(t+\sqrt{t^2-1})^2}
\end{eqnarray}
and 
  $c_0 = - {\pi \sqrt{3}}/{12} + {\ln 3}/{4} + {1}/{4} \approx 0.0712$.

\begin{figure}[h]
\begin{center} \epsfig{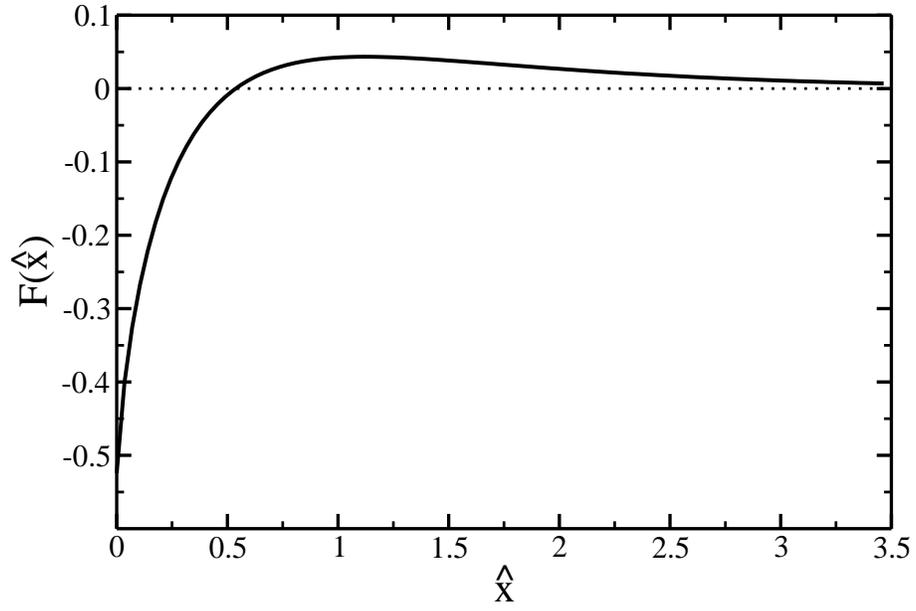}
\end{center}
\caption{$F$ function, proportional to the charge density profile,
as a function of the distance to the wall in reduced units.
}\label{fig:g_q}
\end{figure}

The generic form of the function $F(\hat{x})$ is given in figure \ref{fig:g_q}.
At the wall, we have $F(0)=-2/3 + 2c_0$, thus the charge density at contact is
\begin{eqnarray}\label{eq:qde0}
  q(0)=z_{as}z_{is}\rho\eta\left(c_0-\frac{1}{3}\right)
\end{eqnarray}
We also note that the integral of $F$ is zero implying that as expected
the profile verifies the electroneutrality condition
$\int q(\hat{x}) d\hat{x} =0$.

\section{Potential profile at the PZC}\label{sec:vpzc}

\begin{figure}[h]
\begin{center} \epsfig{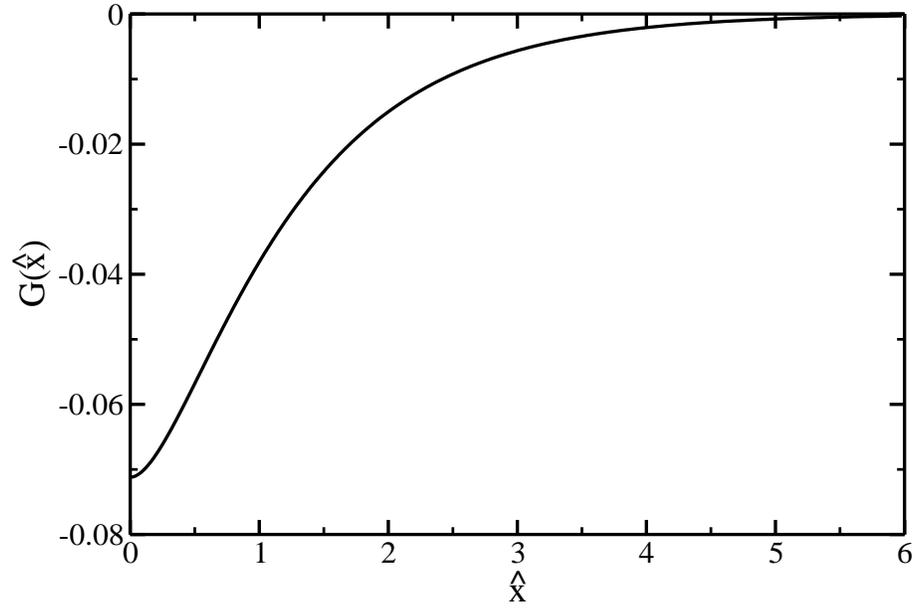}
\end{center}
\caption{$G$ function, proportional to the electric potential profile,
as a function of the distance to the wall in reduced units.
}\label{fig:psiren}
\end{figure}

The electric potential can be obtained using the 
charge profile from the exact expression
\begin{eqnarray}
  \beta e \psi({x}) &=& - \frac{\beta e^2}{\varepsilon}
      \int_{\hat{x}}^\infty ({x}' - {x}) q({x}') d{x}'
\end{eqnarray}
using equation (\ref{eq:g_q}), we obtain
\begin{eqnarray}
  \beta e \psi({x}) &=& 
   - \frac{z_{as}}{z_{is}} \frac{\eta}{2}
      \int_{\hat{x}}^\infty (\hat{x}' - \hat{x}) F(\hat{x}') d\hat{x}'\\
   &=& \frac{z_{as}}{z_{is}}\eta G(\hat{x})\label{eq:vpzc}
\end{eqnarray}
where
\begin{eqnarray}
    G(\hat{x}) &=& \int_1^\infty
    \frac{e^{-2 \hat{x} t}dt}{\left(2t^2+2t\sqrt{t^2-1}-1\right)(4t^2-1)}
     - c_1 e^{-\hat{x}}
\end{eqnarray}
with $c_1 = {\pi \sqrt{3}}/{24} + {\ln 3}/{8} - {1}/{4} \approx 0.11405$.
The generic form of the function $G(\hat{x})$ is given in figure \ref{fig:psiren}.
We note that the derivative of this electric potential, which is proportional
to the electric field, is zero at the interface as expected for
a neutral hard wall.
And at the contact with the wall, we have
\begin{eqnarray}
    \beta e \psi(0) = - \frac{z_{as}}{z_{is}} \eta\;c_0
\end{eqnarray}

\section{Charge contact theorem}\label{sec:chargecontact}
In \cite{chargecontact}, we have derived a charge contact theorem which relates
the contact value of the charge profile to the electric field accross the
interface. Using this exact relation, in the following, we verify the
consistency of our expressions for the charge density and for the electric
potential profiles.
For an asymmetric in valence system of point ions, the charge contact theorem reads
\begin{eqnarray}
  q(0) = \beta e \int_0^\infty (z_+^2\rho_+(x)+z_-^2\rho_-(x))
    \left(\frac{\partial \psi(x)}{\partial x}\right) dx
    + \beta (z_+ P_+ + z_- P_-).
\end{eqnarray}
Note that in comparison to the notations in \cite{chargecontact}, here, $q$ is a
density and does not include the electric charge.
At the lowest order in the loop expansion, we can take the profile of the density
and the electric potential respectively at the zero loop order which corresponds
to a constant profile and at the first order, which is our result equation (\ref{eq:vpzc}).
So the first term becomes
\begin{eqnarray}\label{eq:psi0}
  - \beta e z_{is}^2\rho \psi(0) = (z_+-z_-)\rho\eta c_0 = z_{as}z_{is}\rho\eta c_0
\end{eqnarray}
For the second term, at the same order, the partial pressures are given by the Debye
approximation
\begin{eqnarray}
  \beta P_\pm = \frac{z_\pm \beta P}{z_++z_-}
\end{eqnarray}
where $\beta P = - K_D^3/(24\pi)=-\rho\eta/3$ is the total pressure.
We thus have 
\begin{eqnarray}
  \beta (z_+ P_+ + z_- P_-) &=& \beta (z_+ - z_-)P\\
   &=& -\frac{1}{3} z_{as}z_{is} \eta\rho\label{eq:betaPasym}
\end{eqnarray}
We can see that the sum of the two contributions equations (\ref{eq:psi0}) and
(\ref{eq:betaPasym}) gives the value obtained from the direct calculation of the
charge density profile given in equation (\ref{eq:qde0}).
This shows the consistency of our results for the charge density and electric
potential profiles obtained at the first order in the loop expansion with the
exact contact charge relation.

\section{Discussion}\label{sec:discussion}
In section \ref{sec:g_q} and \ref{sec:vpzc}, we have shown the existence of a
spontaneous polarisation of a neutral interface.
This phenomenon is absent in symmetric systems where we have the same number of
anions and cations at each point accross the neutral interface, although we
have shown there is a profile for the total number of ions \cite{ddcjsjpbElectActa2003}.
The polarisation is then directly related to the asymmetry in valence of the ions,
that is to the deplacement of the equilibrium in number of the ionic species in
order to satisfy the electroneutrality condition.
A way of understanding this profile is to consider the depletion of the
ionic profiles due to the electrostatics existing at interfaces \cite{ddcjsjpbElectActa2003}.
This depletion is stronger and shorter ranged the higher the charge of the ions.
Assuming $z_+>z_-$, for the more highly charged cations, we understand
that their profile is more depleted closer to the interface leading to a
negative charge profile at contact.
This is followed by a distribution of negative ions which is then electrically
compensated by positive ions further away from the wall.
The electric potential at the wall is then negative and increases to reach zero
towards the bulk.
Such qualitative statements have been discussed on the occasion of the contact
theorem for asymmetric electrolytes \cite{chargecontactasym}.

More quantitatively, we have seen in Section \ref{sec:chargecontact}, that the
profiles are such that they verify consistently the charge contact theorem
at given order of approximation.
From equation (\ref{eq:g_q}) and (\ref{eq:vpzc}), the expressions of the charge
density and the electric potential profiles are analytic and simple. Here are
some of their characteristic features.
The two profiles depend respectively on the universal functions $F$ and $G$ in
terms of the reduced distance $\hat{x}$ and linearly on the loop expansion
parameter $\eta$. As a consequence, we can see that similar to the desorption
phenomenon and to the anomalous capacitance behaviour \cite{dungbodaasym}
this polarisation of the interface increases as the reduced temperature
decreases. We recall that the reduced temperature is proportional to the
temperature and to the dielectric constant and inversely proportional to the
product of the ionic charges. Therefore, there is a fair number of physical
systems which correspond to a low reduced temperature as it is possible to
decrease the systems' temperature or dielectric constant or conversely consider
highly charged ionic species.
Typically for a 0.05 molar solution of a 2:1 electrolyte at room temperature
in water we have $\eta\approx 0.909$.
Then, the value of the profiles are: linear on the difference on the
product $z_{as}z_{is} = z_+ - z_-$ for the charge profile where as the reduced electric potential
scales on the ratio $z_{as}/z_{is} =(z_+ - z_-)/(z_+ z_-)$.
As a consequence the behaviour of the charge and of the electric potential are
different. The charge scales proportionally to the difference $z_+-z_-$, where
as the ratio $z_{as}/z_{is}$ is bound by $1$ for high charge asymmetries.\\

\begin{figure}[h]
\begin{center} \epsfig{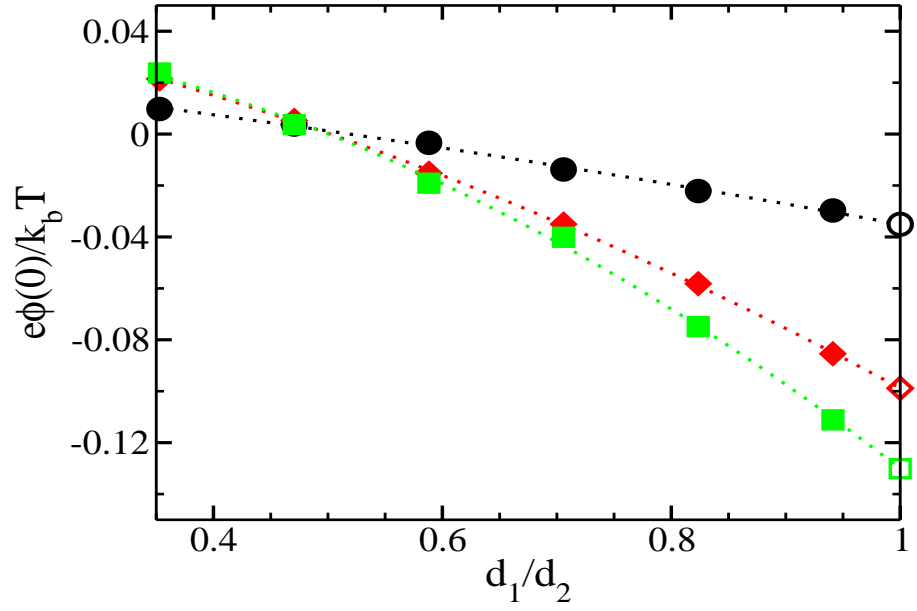}
\end{center}
\caption{PZC data from \cite{HendersonAsym} (circles, diamonds and squares)
for the size asymmetric 2:1 electrolytes for concentrations 0.05, 0.5, 1.0 Mol/l respectively
extrapolated to the equal diameter systems (empty symbols).
The dotted lines are the LLSQ fit used.
}\label{fig:extrapol21}
\end{figure}

\begin{figure}[h]
\begin{center} \epsfig{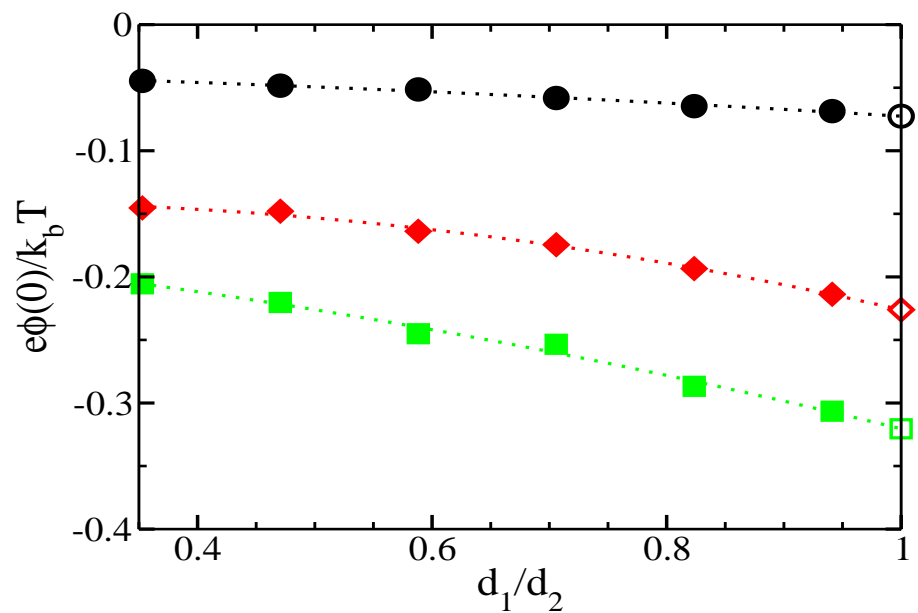}
\end{center}
\caption{Same caption as in figure \ref{fig:extrapol21} for the 3:1 electrolyte
and concentrations 0.033, 0.333, 1.0 Mol/l.
}\label{fig:extrapol31}
\end{figure}

\begin{figure}[h]
\begin{center} \epsfig{file=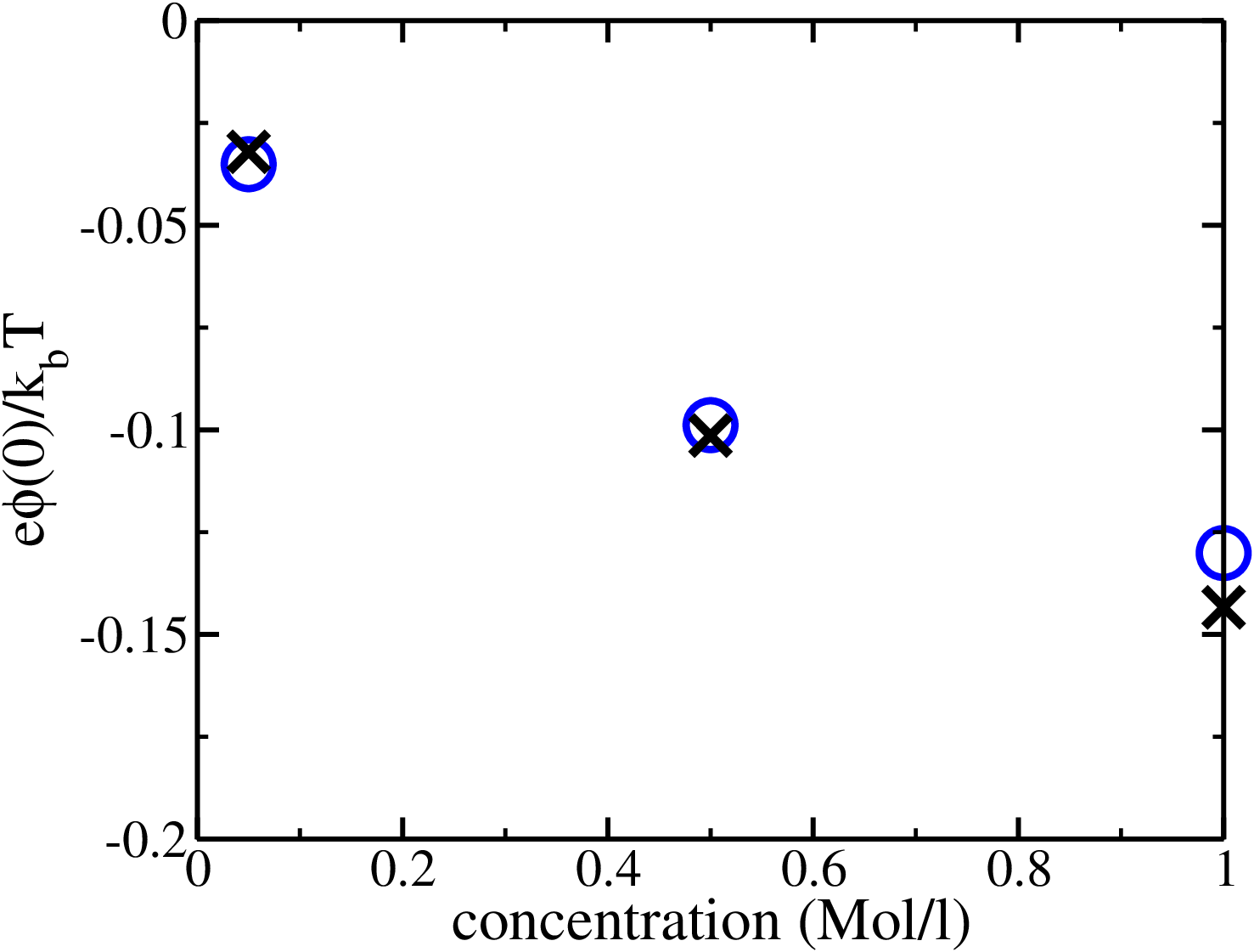,height=8.0cm,width=12.cm}
\end{center}
\caption{PZC for the extrapolated MC simulation results (empty circles) of figure
\ref{fig:extrapol21} and the field theory  Table I (crosses) for a 2:1 electrolyte as a function
of the ionic concentration.
}\label{fig:phi0MCFT21}
\end{figure}

\begin{figure}[h]
\begin{center} \epsfig{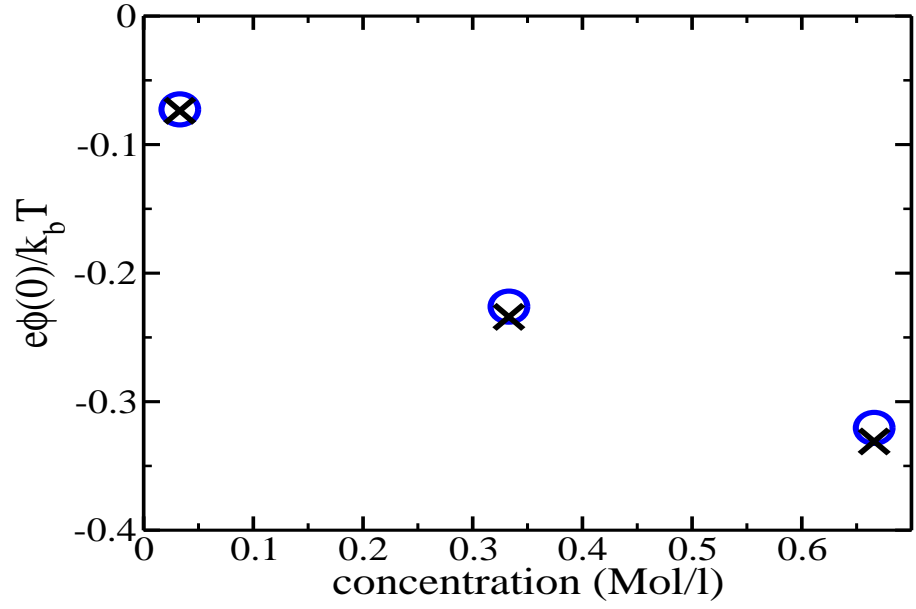}
\end{center}
\caption{PZC for the extrapolated MC simulation results (empty circles) of figure
\ref{fig:extrapol31} and the field theory Table II (crosses) for a 3:1 electrolyte as a function
of the ionic concentration.
}\label{fig:phi0MCFT31}
\end{figure}

We have compared quantitatively our point ion expression for the PZC with the
numerical simulation results in \cite{HendersonAsym}.
Note that for this comparison, the contact value distance corresponds to half
the ionic diameter for the finite sized ions of the MC simulations, where as for
the point ions it corresponds to a vanishing distance to the wall.
As the size asymmetry is immaterial for the point ions, we first extrapolate the
simulation results for the equal size ions using a linear least squares
quadratic (LLSQ) fit for the 2:1 and 3:1 electrolytes as shown in figure
\ref{fig:extrapol21} and \ref{fig:extrapol31}.
The fits give for the various concentrations used in \cite{HendersonAsym}
the PZCs' shown in Table I and II, were  we have also given the value of the parameter $\eta$.
\begin{table}
\caption{Extrapolation of the PZC by LLSQ fit for the 2:1 electrolyte from numerical
data results in \cite{HendersonAsym}.}
\centerline{{\begin{tabular}{|c|c|c|}
\hline
 concentration & PZC & $\eta$ \\
\hline
 0.05 Mol/l \; & \;  -0.0351 \; & \; 0.909\;\\
\hline
 0.5  Mol/l \; & \;  -0.0989 \; & \; 2.87\;\\
\hline
 1.0  Mol/l \; & \;  -0.130 \; & \; 4.07\;\\
\hline
\end{tabular}}}
\end{table}
\begin{table}
\caption{Extrapolation of the PZC by LLSQ fit for the 3:1 electrolyte from numerical
data results in \cite{HendersonAsym}.}
\centerline{{\begin{tabular}{|c|c|c|}
\hline
 concentration & PZC & $\eta$\\
\hline
 0.033 Mol/l \; & \;  -0.0726 \; & \; 1.57\;\\
\hline
 0.333 Mol/l \; & \;  -0.226 \; & \; 4.98\;\\
\hline
 0.666 Mol/l \; & \;  -0.320 \; & \; 7.04\;\\
\hline
\end{tabular}}}
\end{table}
For all these
systems we note that $\eta \geq 1$, which indicates the importance of beyond
mean field correlations.
In contrast to the numerical simulation results, simple, analytic approaches of the
liquid state theory like the LMGC (Linear Modified Gouy Chapman) and MSA (Mean
Spherical Approximation) theory give a vanishing value of the PZC \cite{HendersonAsym}.\\
The comparison of the extrapolated values of the potential with the field theory are given
as a function of the density in figure \ref{fig:phi0MCFT21}
and \ref{fig:phi0MCFT31}.
For nearly all values, the results of the field theory for the simple point ions are
very close to those of the simulations, for the concentrations investigated
and for both 2:1 and 3:1 electrolyte. The only point which is not as accurate
corresponds to the highest density considered for the 2:1 electrolyte.
The overall agreement of the theory seems to indicate, that the excluded volume
effects are not crucial.  And favours of our interpretation that this
polarisation phenomenon is essentially entropic in nature meaning it is related
to the number balance of the ionic species, which follows the valence asymmetry.\\

\begin{figure}[h]
\begin{center} \epsfig{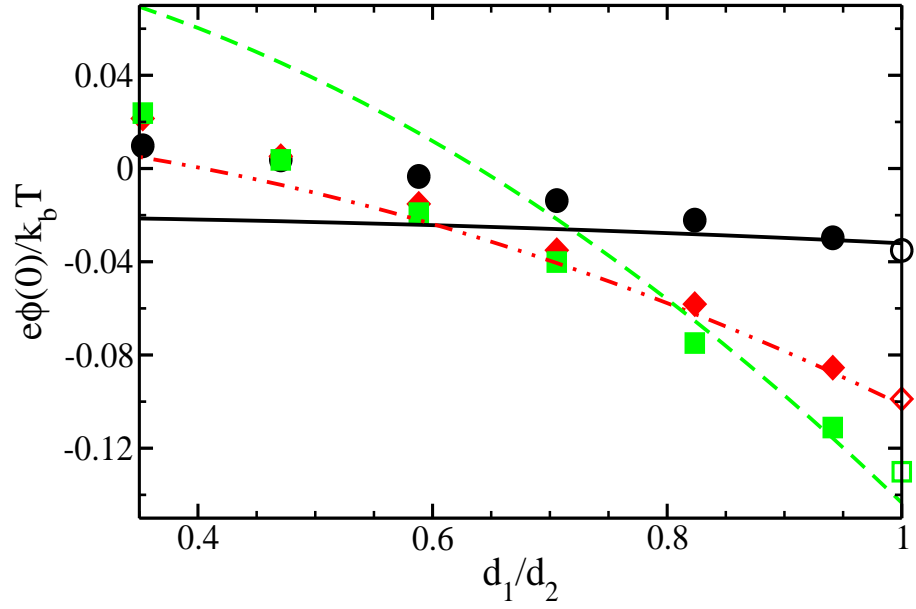}
\end{center}
\caption{PZC as a function of the diameter ratio, for a 2:1 electrolyte and the
same physical parameters temperature, dielectric constant
as in \cite{HendersonAsym}.
The symbols (circles, diamonds and squares) are for the numerical simulations
data and the empty symbol for the extrapolated value for equal diameters, and
the curves (full, dotted-dashed and dashed) are for the field theory plus the LMGC theory
\cite{HendersonAsym} presented in the order of the concentrations
0.05, 0.5, 1.0 Mol/l.
}\label{fig:lmgc21}
\end{figure}

\begin{figure}[h]
\begin{center} \epsfig{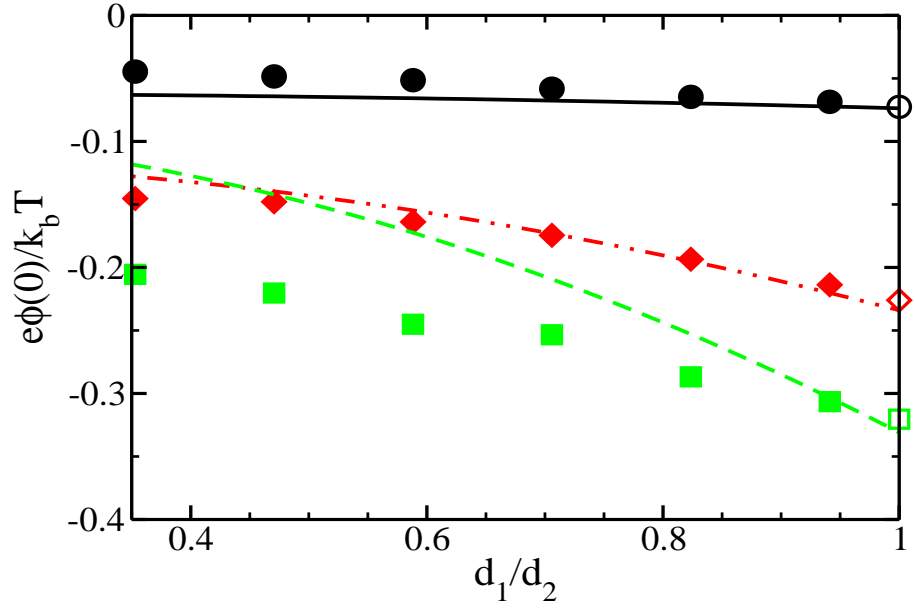}
\end{center}
\caption{PZC as a function of the diameter ratio, for a 3:1 electrolyte and the
same physical parameters temperature, dielectric constant
as in \cite{HendersonAsym}.
The meaning of the symbols and curves are those of figure \ref{fig:lmgc21}
except that the concentrations are 0.033, 0.333, 1.0 Mol/l.
}\label{fig:lmgc31}
\end{figure}

\begin{figure}[h]
\begin{center} \epsfig{file=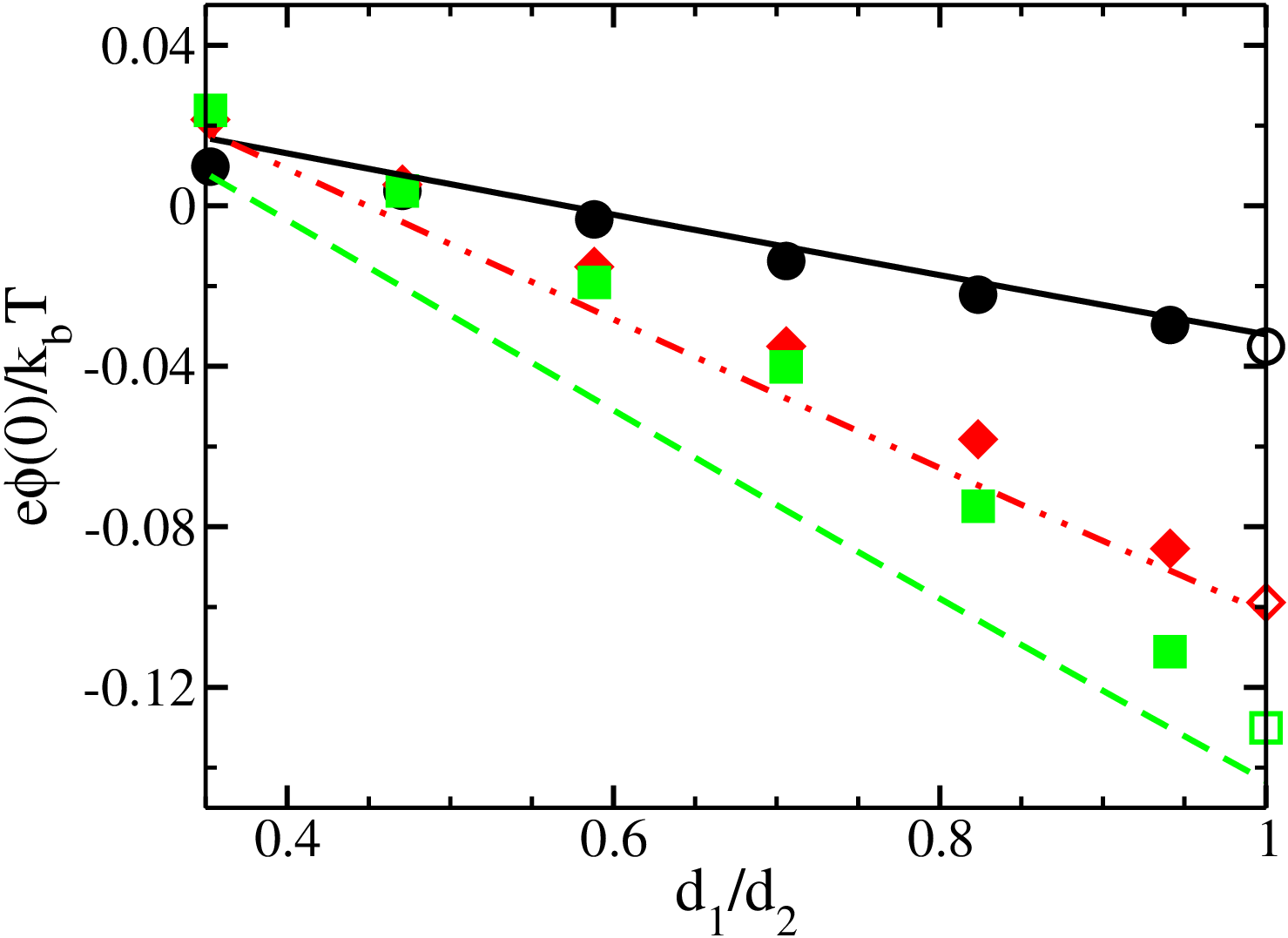,height=8.0cm,width=12.cm}
\end{center}
\caption{PZC as a function of the diameter ratio, for a 2:1 electrolyte,
with caption identical to that in figure \ref{fig:lmgc21} except
that the MSA approximation is used in place of the LMGC approximation.
}\label{fig:msa21}
\end{figure}

\begin{figure}[h]
\begin{center} \epsfig{file=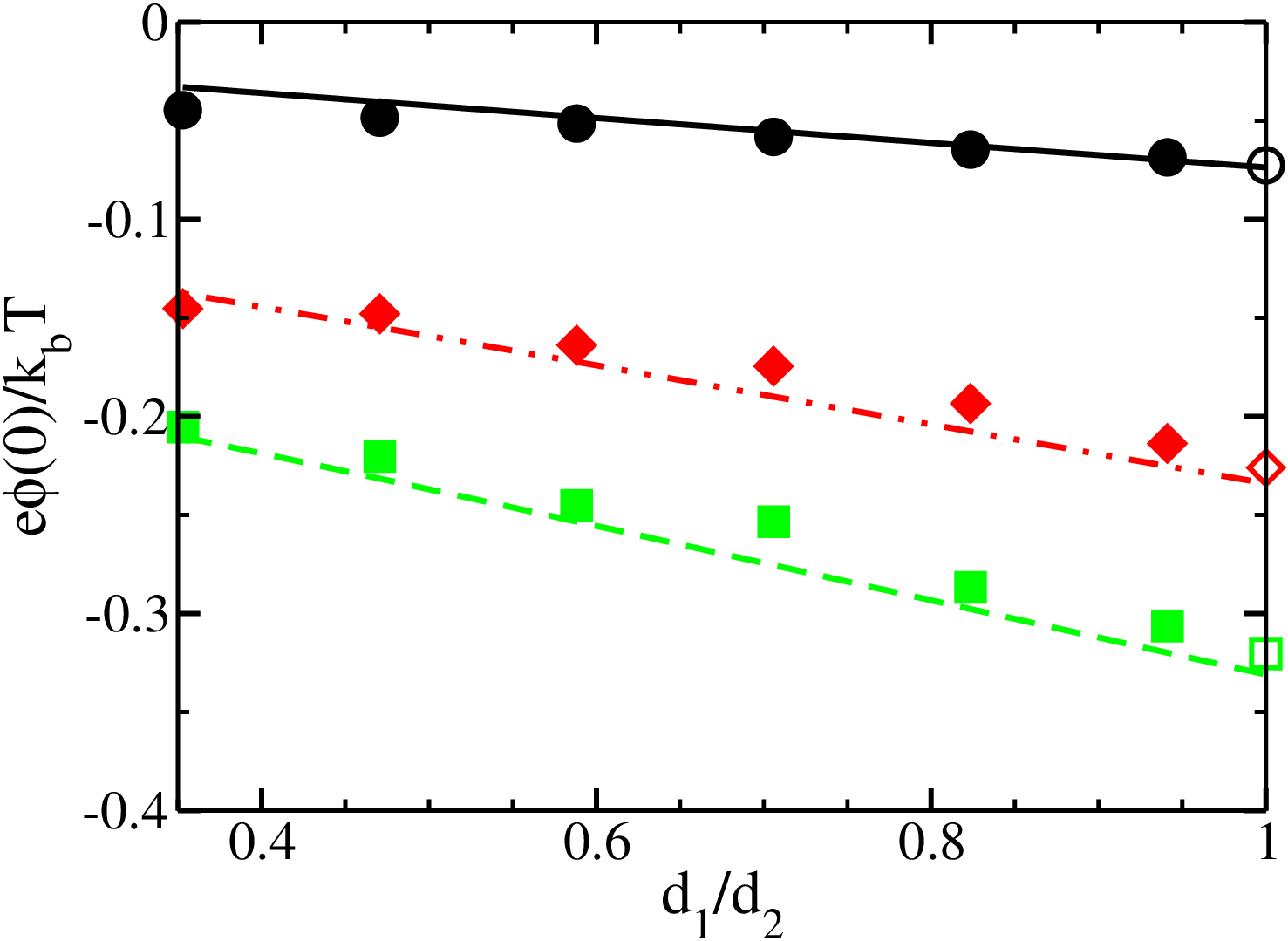,height=8.0cm,width=12.cm}
\end{center}
\caption{PZC as a function of the diameter ratio, for a 3:1 electrolyte,
with caption identical to that in figure \ref{fig:lmgc31} except
that the MSA approximation is used in place of the LMGC approximation.
}\label{fig:msa31}
\end{figure}

Given the results of the simple point ion model, we have considered extending the
comparison for non symmetric electrolyte both in size and in valence as they are
studied in \cite{HendersonAsym}. In this article, for valence symmetric systems
the LMGC and MSA approaches give a reasonable dependence with respect to the
ionic diameter ratio.
It is then tempting to assume an additive approximation combining the size
asymmetry well described by standard liquid state theory approaches and the
valence asymmetry which is well described by our field theory approach.
We have used this assumption for the results presented in figures
\ref{fig:lmgc21} - \ref{fig:msa31}.\\
In figures \ref{fig:lmgc21} and \ref{fig:lmgc31}, the theory is obtained by
adding the PZC values of the LMGC approach to those of the field theory
(Tables I and II), respectively for the 2:1 and 3:1 electrolytes and for the corresponding
concentrations.
For both types of electrolytes, this approximation gives good results
for near symmetric in size systems.
At the intermediate density considered, the agreement extends over
all diameter ratios whereas it respectively underestimates and
overestimates the effect at low and high densities.\\
In figures \ref{fig:msa21} and \ref{fig:msa31}, we replace
the PZC values of the LMGC by those of the MSA approach.
In this case, the results are overall much closer to the simulation results
for both type of valences and all concentrations.
However, in the case of the 2:1 electrolyte the approximation
is more succesful at lower concentration.
Whereas for the 3:1 electrolyte, the comparison is incredibly good over the
whole concentration range considered.

Thus simply adding the PZC effects due to the size and valence
asymmetries gives rather good results and allows with a simple model
the simultaneous description of both asymmetries.
We believe that the discrepancy appears to be lower for the 3:1 electrolyte
as in this case the valence effect which is rather well predicted
by the field theory is larger and possibly may partly hide the discrepancies
in the description of the size effects.

\newpage
\section{Conclusion}
In this paper, we have studied the behaviour of valence asymmetric electrolytes
at simple hard neutral interfaces.
We have discussed the existence of a polarization of the neutral interface for
these systems which is not related to any specific interaction with the interface.

Using a point ion model, we show that this effect is purely entropic. More
precisely, it is not related to volume exclusion potential effects rather it is
connected to the number balance between the ions.
This phenomenon is quite fundamental as its origin comes from the combinatorics
and counting of the states in the phase space set by basic rules of the quantum
mechanics. These familiar rules give the entropic contribution of the particles
of the same kind which are indiscernible and as a consequence also set the
combinatory for anions and cations which are discernible species.
The physics of these rules appears in the entropic part of the functional of our
field theory Hamiltonian.
The role of this entropy has previously been emphasized discussing the
desorption phenomenon and the related anomalous capacitance behaviour. In this
case, the entropic functional leads to a coupling between charge and total
density fields \cite{anomalouscapa,densityboda}.
In the present work, the same entropic functional gives couplings for the
charge field alone and gives the correction to the trivial vanishing mean
field charge density profile.
Another interesting aspect, in the approach, is that on the field
variables we are allowed algebraic operations (e.g. linear combinations of the
fields like the charge or the total density) which turn out to be meaningful
physically and are less straightforward to implement when discussing particles.
Finally, for the charge and the electric potential profile,
we obtain expressions which are analytic and rather simple with well
identified scaling behaviours in terms of the physical parameters.
In particular, we discuss the role of two parameters related to multivalency.
One is the product of the valences, which is related to the
ionic strength in the system.
The second is more characteristic of the asymmetry as it is proportional
to the difference of the valences.

We believe that this study illustrates the interest of our FT approach.
One important aspect is the original description of the quantum mechanical
degrees of freedom which leads to new interpretations and allows for certain
systems to obtain a clear physical picture and simple application in comparison
to other approaches.

\subsection*{Acknowledgment}

M.~Holovko and D.~di~Caprio are grateful to J.P. Badiali for enlightening
discussions and for the support of the National Academy of Science
of Ukraine (NASU) and the CNRS, in the framework of project n\symbol{23}21303.
The authors are also grateful to M. Valisk\`o, D. Boda and D. Henderson for providing
their simulation data results \cite{HendersonAsym}.

\clearpage

\clearpage

\clearpage

\clearpage

\clearpage

\clearpage

\clearpage

\clearpage

\clearpage

\clearpage

\clearpage

\end{document}